# A Study of Thermocurrent Induced Magnetic Fields in ILC Cavities

Anthony C. Crawford     Fermilab     acc52@fnal.gov

The case of axisymmetric ILC type cavities with titanium helium vessels is investigated.
A first order estimate for magnetic field within the SRF current layer is presented.
The induced magnetic field is found to be not more than $1.4 \times 10^{-8}$ Tesla = 0.14 milligauss
for the case of axial symmetry.  Magnetic fields due to symmetry breaking effects are discussed.

## Introduction

It has been reported that electrical current driven by Seebeck effect thermal voltage creates a magnetic field that when trapped in the niobium of a superconducting RF cavity results in a surprisingly large contribution to RF surface resistance [1]. This is due to the cavity-helium vessel assembly consisting of two metals, niobium and titanium, with different Seebeck coefficients ($S$), ie, essentially an arrangement of nb-ti thermocouples. If it is the case that the induced magnetic field is significantly large, then it would be advantageous to lower the temperature of the cavity - helium vessel assembly through $T_c$ (9.2K) in a manner that minimizes any thermal gradient. The practical way to achieve minimal thermal gradients is to cool the cavity very slowly.

Results from several SRF institutions are in disagreement. Helmholtz Zentrum Berlin sees lower surface resistance with slow cooldown. Fermilab, Jlab and DESY see higher surface resistance with slow cooldown. Cornell has seen both increased and decreased surface resistance from slow cooldown.

In order to clarify the potential contribution to surface resistance from thermomagnetic fields it is useful to study the basic geometry of the cavity-helium vessel system and to perform simple calculations relying on previously measured and documented electrothermal material properties for niobium and titanium.

## A Model of the Cavity Geometry

The basis of the model is shown in Figure 1. Note that throughout this report the color green is used for titanium and grey for niobium. Rotational symmetry about the axis of the cavity is assumed.

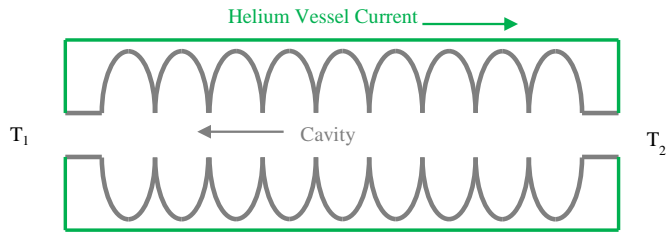
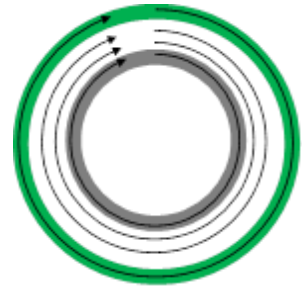

Figure 1.  System Geometry                    Figure 2.  Magnetic Field Lines

A difference in Seebeck coefficient in the niobium compared to the titanium parts of the closed circuit will drive a current if there is a temperature difference along the cavity axis. The cavity-helium vessel system is similar to a coaxial transmission line. The magnetic field that results from the thermal current will be confined to the region between the inner surface of the niobium cavity and the outer surface of the titanium helium vessel. The direction of the magnetic field will be in the direction of the angle θ with respect to the cavity axis. Figure 2 shows the direction and location of magnetic field lines with respect to a view in the direction of the cavity axis.

## Material Properties

In this section relevant thermoelectric material properties will be listed and described.   Thermopower characterization for the system is considered first. The graph shown in Figure 3 was made using data listed in reference [2]. No data for titanium below 50K was available, so the titanium curve was forced to fit a Seebeck Coefficient value of zero at 0K

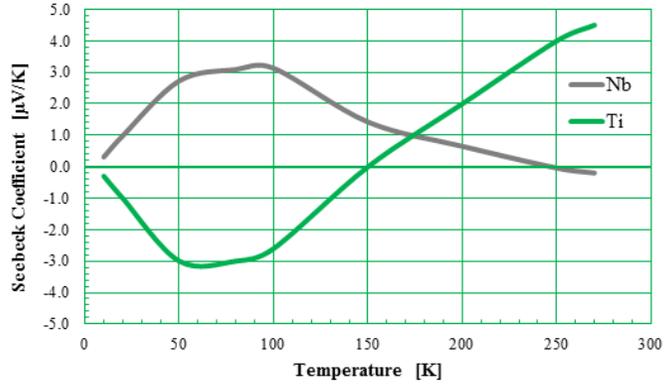

Figure 3.    The Seebeck Coefficient for Niobium and Titanium at Low Temperatures

The electrical resistivity of niobium and titanium at low temperatures is also required. These are shown in Table 1. A residual resistivity ratio [RRR] value for cavity niobium of 300 is typical and is assumed here. The source for the RRR value for titanium is reference [3].

Table 1.    Electrical Resistivity for Niobium and Titanium at Low Temperatures

|  | $\rho_{300}$ [$\Omega$-m] | RRR | $\rho_{10}$ (calculated) [$\Omega$-m] |
|---|---|---|---|
| niobium | $1.52 \times 10^{-7}$ | 300 | $5.1 \times 10^{-10}$ |
| titanium | $4.2 \times 10^{-7}$ | 10 | $4.2 \times 10^{-8}$ |

**A First Order, Worst Case Estimate for Thermocurrent Induced Magnetic Field**

The equivalent electrical circuit for the cavity-helium vessel system can be made particularly simple and is shown in Figure 4. Because the titanium bellows in the helium vessel is thin, a useful underestimate for total circuit resistance is to assume that all of the resistance in the electrical circuit is contributed by the bellows. The resistance of the bellows is calculated from the parameters listed in Table 2.

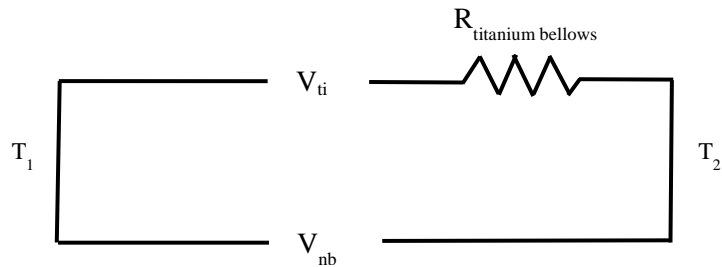

Figure 4.    The Equivalent Circuit

Table 2.    Electrical and Mechanical Properties of the Titanium Helium Vessel Bellows

| | | |
|---|---|---|
| Folded Length | | 0.03 m |
| Unfolded Length | L | ~ 0.09 m |
| Thickness | | 3.05 x $10^{-4}$ m |
| Average Diameter | | 0.23 m |
| Cross Section Area | A | 2.20 x $10^{-4}$ m$^2$ |
| Electrical Resistivity | $\rho$ | 4.2 x $10^{-8}$ $\Omega$-m |
| Resistance | R | 2.20 x $10^{-4}$ $\Omega$ |

The electrical resistivity for titanium listed in Table 2 was chosen to be the resistivity for RRR 10 titanium at 10K listed in Table 1. It is unlikely that the electrical resistivity of the bellows could be significantly less than this value. The resistance is calculated from the definition of electrical resistivity:

$$R = \frac{\rho L}{A}$$

In order to calculate the current in the circuit, it is necessary to estimate the total circuit electromotive force ($\varepsilon$) using the Seebeck Coefficients of niobium and titanium. It is assumed for the purpose of this calculation that one end of the cavity-vessel system is at 9.2K and that the other end is at 175K. Because of the change of sign in the coefficient of titanium at 150K, this is a worst case choice for $T_2$ that leads to the largest possible value of $\varepsilon$. $\varepsilon$ is given by the expression:

$$\varepsilon = \int_{T_1}^{T_2} [S_{nb}(T) - S_{ti}(T)] dT$$

Reasonably good non-phenomenological fits to the Seebeck Coefficient curves for niobium and titanium are given by the fourth order polynomials:

$$S_{nb}(T) = -6.0 \times 10^{-9} T^4 + 5.0 \times 10^{-6} T^3 - 1.4 \times 10^{-3} T^2 + 0.1341 T - 1.0285$$

And

$$S_{ti}(T) = 7.0 \times 10^{-9} T^4 - 6.0 \times 10^{-6} T^3 + 1.7 \times 10^{-3} T^2 - 0.1549 T + 1.238$$

For the fitted curves, T must be in degrees K and S is in µV/K. The numerical value of the definite integral for $\varepsilon$ is:

$$\varepsilon = 6.5 \times 10^{-4} \text{ Volts}$$

The current in the circuit is then given by:

$$I = \varepsilon/R = 37 \text{ Amperes}$$

The electrical power dissipated in the bellows is only 24 milliwatts and is removed by the flow of helium gas during cooldown.

Enough information has now been gathered to allow calculation of the magnetic field in the SRF layer. The integral form of Ampere's law states that:

$$\oint B dl = \mu_0 I_{enclosed}$$

For purposes of this calculation, it is assumed that the penetration depth of the SRF conduction layer is 200 nanometers = 2.0 x $10^{-7}$ meters. The contour of integration is chosen to be the radially outermost boundary of the RF layer. Ampere's law, when combined with the axial symmetry condition, requires that only the current that lies within a radius less than or equal to the outer radius of the RF layer contributes to magnetic field within the RF layer.

In order to continue with worst case conditions, the diameter of the cavity "iris" is used for the contour of integration: D = 0.070 m. Because it has the smallest radius, the iris is the location of the largest thermomagnetic field. Performing the integral along the circular path gives:

$$B = \mu_0 I_{enclosed} / \pi D$$

The change in temperature of the cavity with respect to time is slow enough so that the thermoelectric current can be considered to be steady state, allowing for the assumption that the current distribution within the niobium of the cavity wall is uniform, ie, j(r) is constant within the niobium. The thickness of the cavity wall is 0.0030 m. The fact that the SRF conduction layer is very thin is a very important feature of this system. Using 2.0 x $10^{-7}$ m for the thickness of the SRF conduction layer means that the fractional current of

$$I_{srf} = (2.0 \times 10^{-7} \text{ m} / 3.0 \times 10^{-3} \text{ m}) \; 37 \text{ Amperes} = 2.5 \times 10^{-3} \text{ Amperes} = I_{enclosed}$$

flows at radii within the SRF layer. The magnetic field in the SRF layer is then given by:

$$B = \mu_0 I_{enclosed} / \pi D = 1.4 \times 10^{-8} \text{ Tesla} = 0.14 \times 10^{-3} \text{ gauss}.$$

By comparison, the ambient magnetic field in SRF test cryostats is typically not less than $5 \times 10^{-3}$ gauss when the temperature is 9.2K. The field value of $1.4 \times 10^{-8}$ Tesla for the case of axial symmetry and 37 Amperes has been verified with a finite element model.

## Discussion

The calculated value for thermoelectric induced magnetic field, based on estimates that tend to maximize it wherever possible, does not allow for a contribution to surface resistance of the cavity that is significantly large when compared to trapped flux from other sources. Of course, there are symmetry breaking perturbations, such as mechanical tolerances and the presence of the two-phase helium connection between the helium vessel and the helium header pipe that may lead to increased trapped flux in the cavity.

Preliminary finite element modeling indicates that for the case of a one millimeter offset in the axis of the cavity from the axis of the helium vessel and a thermocurrent of 37 Amperes (non uniform), there is a magnetic field of $7 \times 10^{-7}$ Tesla = $7 \times 10^{-3}$ gauss within the SRF layer. A mechanical error such as this could double the amount of trapped flux in a well shielded cavity.

The presence, or lack of, large mechanical errors seems to be a plausible potential explanation for the non reproducibility of cavity surface resistance. A combination of worst case thermal gradient and a cavity that is not straight could lead to a large increase in residual resistivity. For the case of concentric cavities and helium vessels, more efficient flux exclusion by fast cooldown through 9.2K may dominate, resulting in lower residual resistance.

Note that the result of this report does not contradict the conclusions of the HZB study of thermoelectric magnetic fields in nb-ti laboratory models [4]. It is the particular axisymmetric geometry of the cavity-helium vessel that differentiates the two results.

For the case of horizontal cryomodule cooldown it is judged that Seebeck voltages are not likely to be larger than those used in this estimate due to the introduction of cold helium gas at a location that is intermediate between the ends of the cavity-helium vessel system.

## Conclusion

The estimated value for thermoelectric currents for the case of axial symmetry is not large enough to support the concept that thermocurrents are responsible for surface resistance degradation during SRF cavity cooldown. Mechanical errors leading to non concentricity of the cavity and the helium vessel are a potential source for significantly large thermocurrent induced magnetic fields.